\documentclass{svproc}
\usepackage{url}
\usepackage{amsmath}
\usepackage{amsfonts}
\usepackage{slashed}
\usepackage{float}
\usepackage{graphicx}

\def\beq{\begin{equation}}
\def\eeq{\end{equation}}
\def\bea{\begin{eqnarray}}
\def\eea{\end{eqnarray}}
\begin{document}
\mainmatter              
\title{ Modified TBM and role of a hidden $\mathbb{Z}_2$}
\titlerunning{TBM mixing and $\mathbb{Z}_2$ symmetry}  
%
\author{Rome Samanta\inst{1} \and Mainak Chakraborty\inst{2}
}
\authorrunning{R. Samanta et al.} 
%
\tocauthor{Ivar Ekeland, Roger Temam, Jeffrey Dean, David Grove,
Craig Chambers, Kim B. Bruce, and Elisa Bertino}
\institute{Saha Institute of Nuclear Physics, HBNI, 1/AF Bidhannagar,
  Kolkata 700064, India \\
\email{rome.samanta@saha.ac.in}
\and
Centre of Excellence in Theoretical and Mathematical Sciences\\
SOA University, Khandagiri Square, Bhubaneswar 751030, India\\\email{mainak.chakraborty2@gmail.com}}

\maketitle              

\begin{abstract}
In a residual $\mathbb{Z}_2 \times \mathbb{Z}_2$ symmetry approach, we investigate  minimally perturbed Majorana neutrino mass matrices. Constraint relations among the low energy neutrino parameters are obtained. Baryogenesis  is realized through flavored leptogenesis mechanism with quasi-degenerate right handed (RH) heavy neutrinos.  
\end{abstract}
\section{Introduction}
In the minimally extended Standard Model (SM) with singlet RH neutrino fields $N_{Ri}$, $\mathcal{O}$(eV) neutrino masses are generated through Type-I seesaw mechanism. Relevant Lagrangian for the latter can be written as 
\bea
-\mathcal{L}_{mass}^{\nu,N}=\bar{\nu}_{L\alpha} (m_D)_{\alpha i}{N}_{Ri}+\frac{1}{2}\bar{N^C}_{iR}(M_R)_i \delta _{ij}N_{jR} + {\rm h.c}. \label{r1}
\eea
with $N^C_i=C\bar{N}_i^T$. The effective light neutrino Majorana mass matrix $M_\nu$ which is  then obtained by the standard seesaw formula $M_\nu = -m_DM_R^{-1}m_D^T, \label{r2}$
can be put into a diagonal form as $U^T M_\nu U=M_\nu^d \equiv \rm diag\hspace{1mm}(m_1,m_2,m_3)$
with $m_i$ assumed to be real.
The effective low energy neutrino Majorana mass term that contains this $M_\nu$ comes out as
\bea
-\mathcal{L}_{mass}^\nu= \frac{1}{2}\bar{\nu_{L\alpha}^C} (M_\nu)_{\alpha\beta}\nu_{L\beta} + {\rm h.c}.. \label{r3}
\eea 
Now in the basis where the charged lepton mass matrix $M_\ell$ is diagonal,  $U$ follows the standard parametrization\cite{1}.
 Let us now have look at the latest $3\sigma$ ranges \cite{2} for the relevant neutrino parameters obtained from oscillation data.
solar: $\Delta m_{21}^2\equiv$ $m_2^2-m_1^2$: $(7.02 \textendash 8.09)\times 10^{-5}$ $\rm eV^{2}$,
atmospheric: $|\Delta m_{31}^2|\equiv$ $|m_3^2-m_1^2|$: $(2.32 \textendash 2.59) \times 10^{-3}$ $\rm eV^{2}$,
 $\theta_{12}$: $31.29^o \textendash 35.91^o$,
$\theta_{23}$: $38.3^o \textendash 53.3^o$,
$\theta_{13}$: $7.87^o \textendash 9.11^o$. Finally, thanks to the Planck for the observed upper bound on the sum of the light neutrino masses; $\Sigma_i m_i<$ 0.23 eV.\\

 In Ref.\cite{3} it is argued that any horizontal symmetry of neutrino Majorana mass matrix $M_\nu$ is a residual $\mathbb{Z}_2\times \mathbb{Z}_2$ flavour symmetry. The symmetry generators $G_i$ obey the relation $U^\dagger G_i U=d_i\label{gnrel}$ with i=2, 3, i.e. there are two independent $G_i$ and hence $d_i$. We can choose these two independent $d$ matrices as $d_2={\rm diag} \hspace{1mm}(-1,1,-1)$ and $d_3={\rm diag} \hspace{1mm}(-1,-1,1)$. Thus for a given $U$, one can calculate $G_2$ and $G_3$ corresponding to $d_2$ and $d_3$ respectively. In this work we focus  particularly on the TBM mixing and calculate the corresponding $G_i$ matrices; $G_{1,2}^{TBM}$ and $G_3^{\mu\tau}$. 
 It can be justified theoretically as well as phenomenologically that $G_1^{TBM}$ is the  only symmetry  which is  viable one to exist as the unbroken $\mathbb{Z}_2$ generator in the Lagrangian. In the next section, we present an ephemeral discussion regarding the implementation of $G_1^{TBM}$ and $G_3^{\mu\tau}$  on the neutrino fields. For further insights related to the application of residual symmetry in the neutrino sector, the readers could have a quick look at Ref.\cite{7}.
\section{Breaking of $\mathbb{Z}_2^{\mu\tau}$: perturbation to the TBM mass matrices }
Depending upon the residual symmetries on the neutrino fields and the phenomenological viability of the textures of the mass matrices, we discuss two cases.\\
Case 1.  At the leading order $G_1^{TBM}$ and $G_3^{\mu\tau}$ transform both the neutrino fields $\nu_L$ and $N_R$  as $\nu_L \rightarrow G_i\nu_L$ and $N_R\rightarrow G_i N_R$. Now we choose a perturbation matrix $M_R^{G_1\epsilon}$ which violates $\mu\tau$ interchange in $M_R$ but respect $G_1^{TBM}$. The leading order mass matrices and the perturbation matrix are of forms
\bea
m_D^0=\begin{pmatrix}
b-c-a&a&-a\\
a&b&c\\
-a&c&b
\end{pmatrix}, \hspace{1mm} M_R^0=\begin{pmatrix}
y&0&0\\
0&y&0\\
0&0&y
\end{pmatrix},\hspace{1mm}
 M_R^{G_1\epsilon} =
\begin{pmatrix}
0 & \epsilon & \epsilon^\prime\\
\epsilon & \epsilon_4 & 0\\
\epsilon ^\prime & 0 & \epsilon_6
\end{pmatrix}
\eea
where $\epsilon=\frac{1}{4}(3\epsilon_4+\epsilon_6)$ and $\epsilon^\prime = -\frac{1}{4}(3\epsilon_6+\epsilon_4)$. Now the effective $M_\nu$ which is invariant under $G_1^{TBM}$ is written as $M_{\nu 1}^{G_{1}^{TBM}}=-m_D^0 M_R^{-1} (m_D^0)^T$ with $M_R=M_R^0+M_R^{G_1\epsilon}$. Since $G_1^{TBM}$ invariance of the effective $M_\nu$ always fixes the first column of the mixing matrix to $(\sqrt{\frac{2}{3}},-\sqrt{\frac{1}{6}},\sqrt{\frac{1}{6}})^T$ up to some phases, a direct comparison of the latter with the $U_{PMNS}$ matrix leads to  a constraint relation between $\theta_{12}$ and $\theta_{13}$ as
 \bea
\sin^2 \theta_{12}=\frac{1}{3}(1-2\tan^2 \theta_{13}). \label{r5}
 \eea
Case 2. In this case, at the leading order, all the neutrino fields obey $G_3^{\mu\tau}$. However  $G_1^{TBM}$ of the over all $M_\nu$ is ensured only by the  transformation $\nu_L\rightarrow G_1^{TBM}\nu_L$. Since the RH singlets are free from $G_1^{TBM}$, the perturbation matrix which is added with $M_R$ is now arbitrary.  Now the most general Dirac mass matrix $m_D^0$, the Majorana mass matrix $M_R^0$ and the perturbation matrix are of the forms
\bea
m_D^0=\begin{pmatrix}
a&\frac{1}{2}(b-c)&\frac{1}{2}(c-b)\\
a&b&c\\
-a&c&b
\end{pmatrix}, \hspace{2mm} M_R^0=\begin{pmatrix}
x&0&0\\
0&y&0\\
0&0&y
\end{pmatrix}, \hspace{2mm}
M_R^{\epsilon}=\begin{pmatrix}
0&0&0\\
0&\epsilon_4 & 0\\
0&0&\epsilon_6
\end{pmatrix}. \label{r6}
\eea
Again the effective $M_\nu$ is calculated as $M_{\nu 2}^{G_1^{TBM}}=-m_D^0 M_R^{-1} (m_D^0)^T $
with $M_R=M_R^0+M_R^{\epsilon}$. Besides reproducing the same relation as obtained in Eq. \ref{r5}, another interesting point  is realized  that $m_D^0$ of (\ref{r6}) is of determinant zero due to the residual $G_1^{TBM}$ symmetry; thus the $M_{\nu 2}^{G_1^{TBM}}$ matrix has one zero eigenvalue. For the remnant $G_1^{TBM}$ symmetry, $m_1$ is of vanishing value.

\begin{figure}[H]
\begin{center}
\includegraphics[scale=.35]{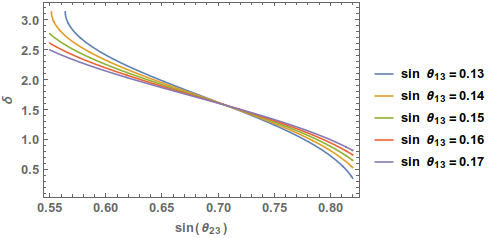}\includegraphics[scale=.35]{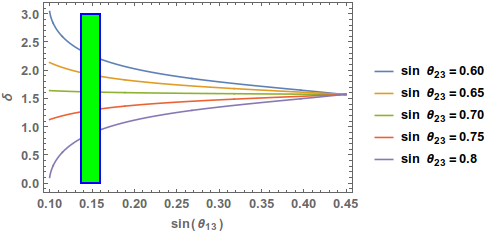}
\caption{Plot in the left side: Variation of $\delta$ with $\theta_{23}$ for different values of $\theta_{13}$. Plot in 
the right side: Variation of $\delta$ with $\theta_{13}$ for different values of $\theta_{23}$ where the green band 
represents the latest $3\sigma$ range for $\theta_{13}$. }\label{dvth}
\end{center}
\end{figure}
 One can also obtain correlation of $\delta$ with the mixing angles\cite{8}.
\section{Flavored  leptogenesis with quasi degenerate RH neutrinos}
 Lepton number, CP violating and out of equilibrium  decays  of RH neutrinos create a lepton asymmetry\cite{4}. A general expression for the CP asymmetry parameter $\epsilon_i^\alpha$ for any RH mass spectrum is given by\cite{5}
\bea
 \frac{1}{4\pi v^2 \mathcal{H}_{ii}}\Big[\sum\limits_{j \neq i} g(x_{ij})\hspace{1mm} {\rm Im}\hspace{1mm}\mathcal{H}_{ij}(m_D)^\dagger_{i \alpha}(m_D)_{ \alpha j}+\sum_{j\ne i}\frac{r {\rm Im}\hspace{1mm}{\mathcal{H}}_{ji}(m_D)^\dagger_{i \alpha }({m_D})_{  \alpha j}}{r^2+\frac{\mathcal{H}_{jj}}{16\pi^2 v^4}}\Big].
\label{epsi_intro_h}
\eea
In (\ref{epsi_intro_h}), $r=(1-x_{ij})$, $\mathcal{H}\equiv {m_D}^\dagger  m_D$, 
$x_{ij}=M_j/M_i$ and $g(x_{ij})$ is given by
\bea
 g(x_{ij})&=&\frac{\sqrt{x_{ij}}(1-x_{ij})}{(1-x_{ij})^2+\frac{\mathcal{H}_{jj}}{16\pi^2 v^4}}+f(x_{ij}).\label{loop}
 \eea
The term proportional to $f(x_{ij})$ comes from the one loop vertex contribution while the remaining are from self energy diagram. Note that in the limit where the RH neutrinos are exactly degenerate, i.e., $x_{ij}=1$, the self energy contribution vanishes and thus a nonzero value of CP asymmetry parameter $\varepsilon^\alpha_i$ is produced only through the vertex contribution. In our model RH neutrinos are quasi degenerate and thereby enhances  the CP asymmetry parameter significantly through  self energy contributions. \\

\begin{center}
\begin{figure}[H]
\includegraphics[scale=0.2]{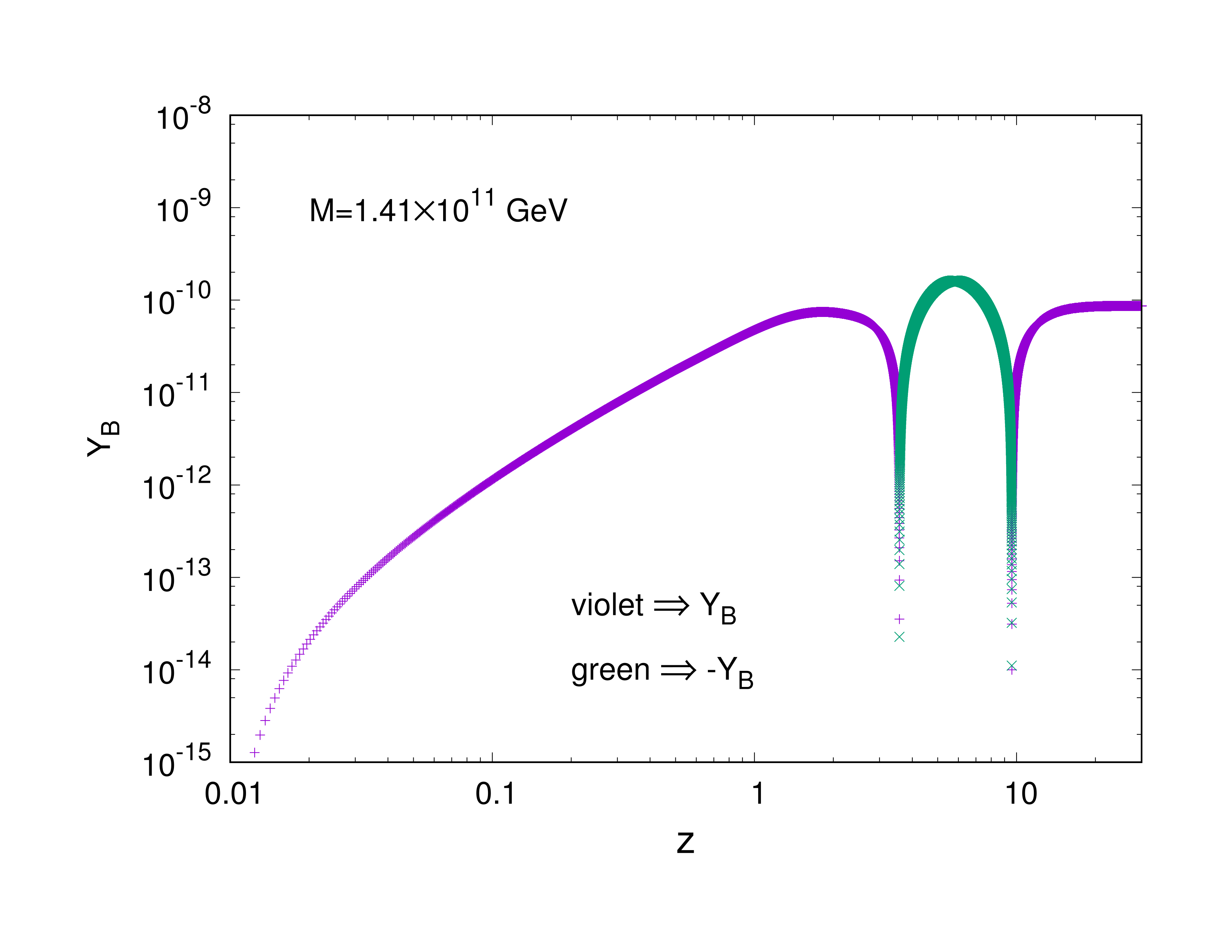}\includegraphics[scale=0.2]{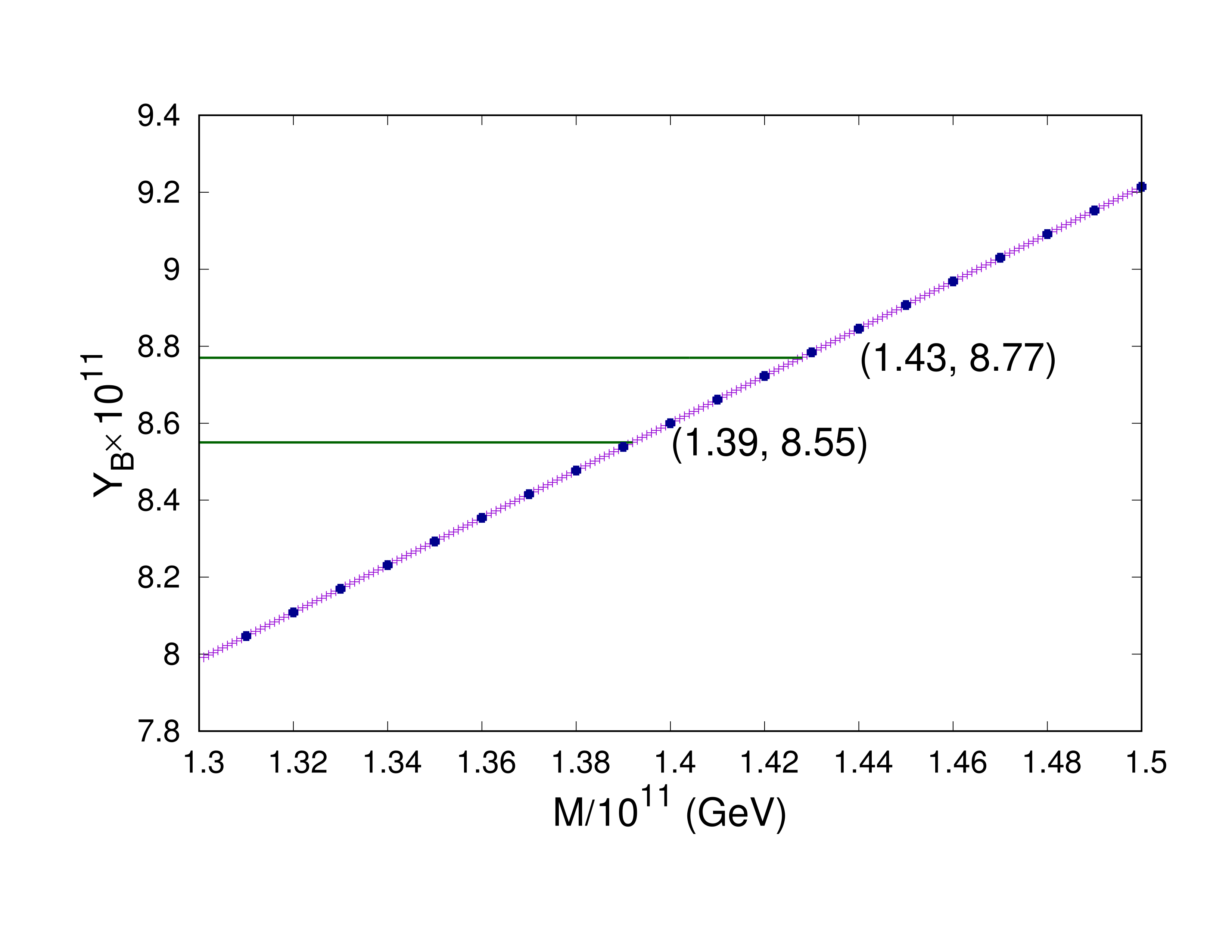}
\caption{Variation of  $Y_B$  with $z$ in the $\tau$-flavored mass regime (left). Upper and lower bounds on $M_1$ for the minimal values of the breaking parameters (right). A normal mass ordering for the light neutrinos has been assumed.}
\label{asy_m}
\end{figure}
\end{center}

Another important issue is that the flavor effect\cite{6} to the produced lepton asymmetry. In \cite{8} we address this issue  in  detail, theoretically as well as  numerically. See Fig.\ref{asy_m} for a typical variation of $Y_B$ with $z=M_1/T$.\\

%
%

\end{document}